\documentclass[twocolumn,english]{article}
\usepackage[T1]{fontenc}
\usepackage[latin9]{inputenc}
\usepackage[a4paper]{geometry}
\usepackage{amsmath}
\usepackage{amssymb}
\usepackage{graphicx}
\usepackage{esint}

\makeatletter
\@ifundefined{date}{}{\date{}}
\usepackage{cite}

\@ifundefined{showcaptionsetup}{}{%
 \PassOptionsToPackage{caption=false}{subfig}}
\usepackage{subfig}
\makeatother

\usepackage{babel}
\begin{document}
\twocolumn[   \begin{@twocolumnfalse}

\title{\textbf{\large Controlling Stray Electric Fields on an Atom Chip for Rydberg Experiments}}
\maketitle
\begin{center}\author{D. Davtyan, S. Machluf, M.L. Soudijn, J.B. Naber, N.J. van Druten, H.B. van Linden van den Heuvell, and R.J.C. Spreeuw }\end{center}
\begin{center}
\textit{Van der Waals-Zeeman Institute, University of Amsterdam, Science Park 904, 1098 XH Amsterdam, the Netherlands} \end{center}
\renewcommand{\abstractname}{\vspace{-\baselineskip}}
\begin{abstract}
Experiments handling Rydberg atoms near surfaces must necessarily deal with the high sensitivity of Rydberg atoms to (stray) electric fields that typically emanate from adsorbates on the surface. We demonstrate a method to modify and reduce the stray electric field by changing the adsorbates distribution. We use one of the Rydberg excitation lasers to locally affect the adsorbed dipole distribution. By adjusting the averaged exposure time we change the strength (with the minimal value less than \(0.2\,\textrm{V/cm} \) at $78\,\mu\textrm{m}$ from the chip) and even the sign of the perpendicular field component. This technique is a useful tool for experiments handling Ryberg atoms near surfaces, including atom chips.
\end{abstract}
\[ \]

\end{@twocolumnfalse}
]

\subsubsection*{\[ \textrm{I. Introduction}\]}

Due to their extreme properties, atoms in a Rydberg state are interesting
objects for creating strongly interacting quantum systems \cite{Saffman2010,Bernien,Saffman2016,Heidemann2007}.
For example, Rydberg atoms strongly interact over large interatomic
distances \cite{Zeiher2016,Labuhn2016,Singer2005a,VanDitzhuijzen2008}
with a van der Waals interaction that scales with principle quantum
number $n$ as $\varpropto n^{11}$ \cite{Ghallagher1994}. This interaction
can be switched on and off by exciting and de-exciting the atoms to
and from the Rydberg state. In addition, Rydberg states have long
life times, $\varpropto n^{3}$, required for quantum information
purposes. However, the large electron orbit also leads to a large
polarizability that scales as $\varpropto n^{7}$, and makes Rydberg
atoms sensitive to electric fields. Managing electric fields is therefore
an important issue for experiments handling Rydberg atoms near the
surface of an atom chip \cite{Teixeira2014,Cisternas2017,Naber2015a}.

An attractive way to realize a scalable Rydberg quantum platform,
is to trap small atomic clouds in arrays of magnetic microtraps close
enough to each other, such that they can interact \cite{Leung2014},
using an atom chip \cite{Leung2011}. However a small intertrap separation
($\sim5-10\,\textrm{\ensuremath{\mu}m}$) also implies a similarly
small distance to the surface. At such a short range the stray electric
fields can be prohibitively large. In fact Rydberg atoms are used
to measure electric fields \cite{Abel2011,Sedlacek2016}.

In this paper we investigate a method to control the electric field
by locally affecting the surface of the atom chip with a blue laser.
The mechanism behind this change is likely a combination of thermally
activated desorption and light-induced atomic desorption (LIAD) \cite{Meucci1994}.
Both these effects are known and used in the cold atoms community.
Changing the temperature of the surface is used to decrease the stray
electric field \cite{Mcguirk2004,Obrecht2007} and thus facilitate
Rydberg excitation \cite{Sedlacek2016}. LIAD has been used by many
groups mostly as a controllable source of atoms and to increase the
number of atoms collected in a magneto-optical trap (MOT) \cite{Burchianti2006,Klempt2006}. 

In this work we focus our Rydberg excitation lasers onto the surface.
This changes the adsorbate distribution on the surface depending on
the duty cycle of one of the lasers. We probe the electric field at
various distances from the chip using two-photon Rydberg spectroscopy.
The measurements show that this technique can affect not only the
strength of the field but also its direction. We model the ad-atom
distribution and the resulting stray field, based on the deposition
and desorption of ad-atom patches during the experimental cycle. We
also take into account the calculated temperature profile caused by
the heating laser. 

This technique holds promise for spatially controlling electric fields
in Rydberg experiments, which is necessary for building a scalable
platform for quantum information with long coherence times.

\subsubsection*{II. Experimental apparatus and measurement technique}

\begin{figure*}[t]
\begin{centering}
\includegraphics[width=0.9\textwidth]{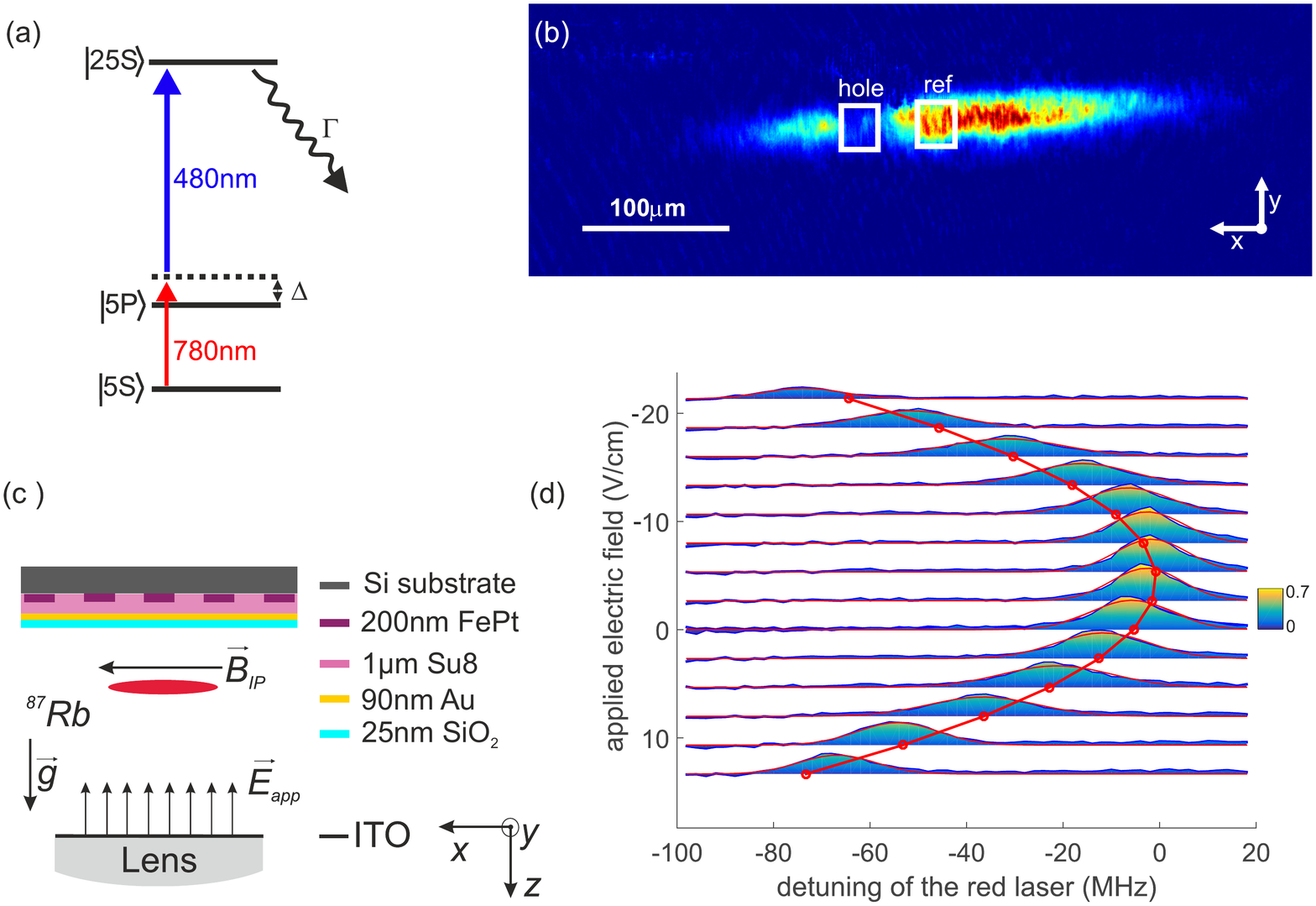}
\par\end{centering}

{\footnotesize{}Figure 1. (a) Two-photon excitation scheme to the
Rydberg state. (b) Example of an absorption image of an atomic cloud,
trapped in a magnetic 'z-wire' trap. The normalized depletion is obtained
by taking the ratio of the number of atoms in the 'hole' area (where
the excitation lasers are focused) and the number of atoms in the
'ref' area. (c) Sketch of the setup (not to scale). The high numerical
aperture lens is coated with ITO which allows the application of a
voltage between the lens and the chip to compensate for the }\textit{\footnotesize{}z}{\footnotesize{}
component of the stray electric field. In order to probe the electric
field on different heights, the atomic cloud (red) can be moved up
and down by varying the current through the trapping wire (z-wire;
not shown). The chip contains a stack of layers of different materials,
including a $1\,\textrm{\ensuremath{\mu}m}$ SU8 polymer layer which
provides thermal insulation and leads to heating of the surface (see
}\textit{\footnotesize{}Temperature simulation}{\footnotesize{}).
(d) An example of a Stark map at a height of $163\,\textrm{\ensuremath{\mu}m}$.
For each value of the voltage between the chip and the lens, a spectrum
is taken by scanning the detuning of the red laser and measuring the
normalized depletion. The red line is a fit to the Stark map; the
values of $E_{z}$ and $E_{x,y}$ are retrieved from the fit.}{\footnotesize \par}
\end{figure*}

Our setup for Rydberg experiments on an atom chip has been described
earlier \cite{Naber2015a}. Briefly, we transfer atoms from a magneto-optical
trap into a z-wire magnetic trap, yielding a cigar-shaped cloud of
$^{87}\textrm{Rb}$ atoms with peak density of $\approx0.14\times10^{12}\,\mathrm{cm^{-3}}$
and a temperature of $\sim3\,\mu\mathrm{K}$. We excite atoms into
the $25S_{1/2}$ state using a two-photon transition. The ground state
$|5S_{1/2},$$F=2,m_{F}=2\rangle$ is coupled to the Rydberg state
through the intermediate state $|5P_{3/2},$$F'=3\rangle$ with a
large blue intermediate-state detuning $\Delta=2\pi\times1.5\,\mathrm{GHz}$
to prevent population of the intermediate state. In Figure 1(a) we
show the level scheme of the two-photon Rydberg excitation using two
lasers: a red laser ($780\,\textrm{nm}$) with a power $P_{\textrm{r}}=35\,\mu\mathrm{W}$,
and a blue laser ($\sim480\,\textrm{nm}$) with a power of $P_{b}=55\,\mathrm{mW}$.
Both laser beams have waists ($1/e^{2}$ radius) of $w_{0}\approx100\,\mu\mathrm{m}$.
This gives a two-photon Rabi frequency $\Omega=\frac{\Omega_{\textrm{r}}\Omega_{\textrm{b}}}{2\Delta}\approx2\pi\times276\,\mathrm{kHz}$.
The one-photon Rabi frequencies ($\Omega_{\textrm{r}}$, $\Omega_{\textrm{b}}$)
were calculated using the Alkali Rydberg Calculator software \cite{Sibalic2017}.

\begin{figure*}[tp]
\begin{centering}
\includegraphics[width=0.95\textwidth]{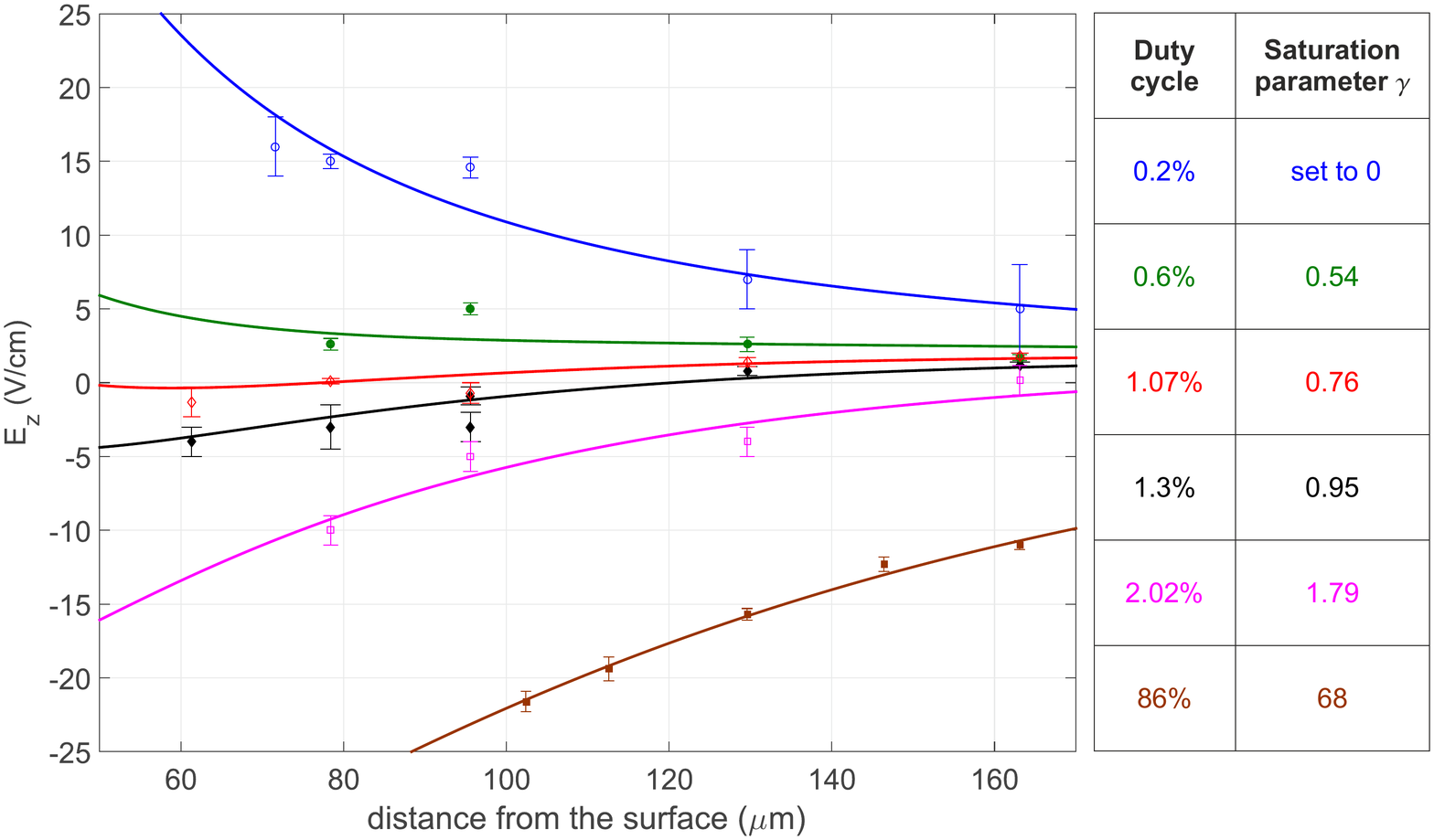}
\par\end{centering}

{\footnotesize{}Figure 2. Dependence of the electric field on the
distance from the chip for different duty cycles of the blue laser.
The result of the measurement of $E_{z}$ is shown with error bars.
The solid lines show the result of a single collective fit (see text
for details). }{\footnotesize \par}
\end{figure*}

In Figure 1(b) we show an example of an absorption image of an atomic
cloud in the z-wire trap. When excited into a Rydberg state, atoms
have a high probability to either decay into an untrappable state
or to be ionized and escape from the trap. Thus we observe a decrease
in the number of trapped atoms in the area where the two laser beams
are focused. The exposure time of the blue excitation laser was varied
between $35\,\textrm{ms}$ to $18\,\textrm{s}$, the exposure time
of the red excitation laser was kept fixed at $100\,\textrm{\ensuremath{\mu}s}$
for all the measurements. Imaging of the cloud was done in-trap $100\,\textrm{\ensuremath{\mu}s}$
after the depletion pulse, so atoms from the neighboring areas don't
have time to refill the depleted area.

In Figure 1(c) we sketch the local environment of the atomic cloud.
The in-vacuum lens ($f=18.75\,\textrm{mm},$ $\textrm{NA}=0.4$),
through which the Rydberg and absorption imaging lasers are focused,
is coated with a conductive material, indium tin oxide (ITO), so that
an electric field $E_{\textrm{app}}$ between the lens and the chip
can be applied and the z-component of the stray electric field ($E_{\textrm{z}}$)
can be compensated. By focusing the two lasers we excite part of the
atomic cloud and observe a depletion because of the Rydberg atoms
escaping the trap. We then scan the frequency of the red laser and
record the loss of atoms. We take spectra for different voltages applied
between the chip and the lens. This results in Stark maps such as
shown in Figure 1(d). The observed peaks are symmetrically broadened,
this can be partially a result of resonant dipole-dipole interaction
between the excited $25S$ state and the neighboring $P$ states,
to which the atoms can decay through black body radiation \cite{Goldschmidt2016,Afrousheh2004,VanDitzhuijzen2008}.
Following \cite{Goldschmidt2016} we estimate our linewidth broadening
to be $\approx2\pi\times3\,\textrm{MHz}$ while the minimum measured
line width is $\approx2\pi\times6\,\textrm{MHz}$, this is in agreement
with the approximate theory. Closer to the surface of the chip the
spectral widths increase. This could be partly due to an increase
of density as the cloud is compressed closer to the surface. Another
possible contribution to the linewidth broadening could be increased
electric field gradients. 

The Stark map has a parabolic shape, as expected for a Ryberg $S$
state for small fields. In the center of the parabola the z-component
of the stray electric field is compensated by the applied field. Thus
the applied compensation field yields a measurement of the stray field
z-component. The zero-frequency detuning corresponds to the resonance,
as calibrated in a spectroscopy vapor cell at room temperature in
a field free environment using electromagnetically induced transparency
(EIT) \cite{Boller1991,Mohapatra2007}. Thus the remaining frequency
shift of the vertex of the parabola is due to the uncompensated $E_{x,y}$
(parallel to the chip surface) component of the stray electric field.
Rydberg energy levels are calculated by diagonalizing the Stark Hamiltonian
using $E_{\textrm{app}}$ as an input parameter and using $(E_{x,y},\,E_{z})$
as fitting parameters. To obtain a good fit, we also introduce as
an extra fitting parameter a factor that enhances $E_{\textrm{app}}-E_{z}$.
This is necessary because some of the measured Stark maps have different
(typically larger) curvature than what is expected from the polarizability
of the investigated Rydberg state. In some measurements we also observe
an asymmetry between the two branches of the parabola which can also
increase the apparent parabola curvature, in a few cases up to a factor
of 4. We verified numerically that this asymmetry does not affect
the position of the vertex to within the error bars. We speculate
that these effects may be due to spatially dependent deposition of
ions or electrons, originating from ionized Rydberg atoms. On the
surface these charges form dipoles with their mirror charge in the
gold layer, and contribute to the stray field. A full understanding
of this complex interplay is beyond the scope of this paper. 

Finally, by changing the current in the z-wire we vary the distance
between the atomic cloud and the atom chip and retrieve the dependence
of the electric field $E_{z}$ on the distance to the chip.

\subsubsection*{\[ \textrm{III. Experimental results}\]}

In previous experiments we observed large stray electric fields above
a gold surface \cite{Tauschinsky2010} and even $\sim10$ times larger
fields above silica-coated gold \cite{Naber2015a}. However, in other
studies \cite{Sedlacek2016} significant reduction of the stray electric
field was achieved by forced desorption of the ad-atoms. In our experiment
we use the focused blue excitation laser to induce local rubidium
desorption through LIAD and heating of the chip surface. We vary the
average laser power, incident on the chip, by the fractional time
(duty cycle) of the blue laser during the experimental cycle. The
pulse length of the red excitation laser ($100\,\textrm{\ensuremath{\mu}s}$)
is kept constant for all the measurements. We then measure the z-component
of the electric field for different duty cycles of the blue laser.
Because the power and the exposure time of the red Rydberg laser and
the imaging laser are much lower than for the blue one, their effect
on the chip surface is negligible. The result of the measurement of
$E_{z}$ on different heights for different duty cycles is shown in
Figure 2. Each of the data sets is taken in a steady state reached
$\sim2.5-3$ hours after changing the duty cycle. The error bars are
standard errors for the fits of the measured Stark maps, the solid
lines show the results of the fitted electric field $E_{z}$ using
the model described below.

The experimental cycle time is $\sim21\,\mathrm{s}$. The uppermost
blue data shows the determined $E_{z}$ for the shortest exposure
time of the blue laser of $\sim35\,\mathrm{ms}$, and thus $0.2\%$
duty cycle. The electric field increases closer to the surface of
the chip. However for higher duty cycles $E_{z}$ decreases and at
some point even changes sign. Below we present a model to describe
this behavior.

\subsubsection*{\[ \textrm{IV. Model} \]}

\subsubsection*{\[ \textrm{\textmd{\textit{a. Electric field simulation}}} \]}

To explain our results we use a model based on a distribution of electric
dipoles, formed by Rb adsorbates on the silica surface. An example
is shown in Figure 3, where the chip (gold and silica layers) is shown
with a dipole distribution (red curve). The positive direction of
the z axis coincides with gravity. The value of $E_{z}$ is indicated
by color intensity. Blue (red) indicates $E_{z}$ pointing down (up).
White color means absence of $E_{z}$. First we assume the distribution
{[}see Figure 3(a){]} is a double Gaussian so that the electric field
points down everywhere. The narrow Gaussian corresponds to atoms released
from the z-wire trap. The wide Gaussian distribution represents atoms
released from the MOT. Once the blue laser is applied to the center
of the dipole distribution for a long enough time, it causes local
desorption of the dipoles. The resulting distribution is shown in
Figure 3(b). It looks like the original distribution with a hole in
the center. In the new case the field lines change direction from
down to up in the center of the hole, and a small region with no stray
field appears (the white area). 

Experimentally, after the blue laser starts to act on the surface,
the desorption of ad-atoms is locally enhanced. We attribute this
effect to a combination of LIAD and a local increase of the temperature.
The temperature of the spot where the beam hits the surface increases
instantly (on the experimental time scale) and accelerates the desorption
of the ad-atoms. When the blue laser pulse ends, the temperature of
the heated area instantly cools down to room temperature. The density
of the remaining ad-atoms (and thus the dipole distribution) depends
on the exposure time of the blue laser during the experimental cycle
(duty cycle).

The dependence of the electric field on the dipole distribution was
simulated using MATLAB software. The z-component of the electric field
at a distance $z$ from the chip produced by a single dipole on the
surface is 

\begin{eqnarray}
E_{z}^{\textrm{s}}(0,0,z) & = & \frac{P}{4\pi\epsilon_{0}}\frac{1}{(x^{2}+y^{2}+z^{2})^{3/2}}\times\nonumber \\
 &  & \left[3\frac{z^{2}}{x^{2}+y^{2}+z^{2}}-1\right],
\end{eqnarray}

\noindent where $(x,y,0)$ are the coordinates of the dipole, $P$
is the dipole moment of a single Rb ad-atom. For $P$ we take the
value $P=12\,\textrm{D}$ \cite{Sedlacek2016,Naber2015a} multiplied
by a correction factor $\frac{\epsilon+1}{\epsilon}\approx1.25$ ($\epsilon=3.9$
is the dielectric constant of silica) to account for image dipoles
due to the gold surface \cite{Obrecht2007,Mcguirk2004,Barrera1978,Pont2015}.
The z axis coincides with the Rydberg excitation beams. The electric
field $E_{z}$ at the point $(0,0,z)$ is then calculated as:

\begin{equation}
E_{z}(0,0,z)=\iintop_{S_{\textrm{ch}}}E_{z}^{\textrm{s}}(0,0,z)\,\rho_{\textrm{d}}(x,y)\,\mathrm{d}x\,\mathrm{d}y,
\end{equation}

\noindent where $\rho_{\textrm{d}}$ is the surface dipole density.
Equation (2) is integrated over the area of the chip $S_{\textrm{ch}}=L_{x}L_{y}$,
with $L_{x}=16\,\textrm{mm}$, $L_{y}=20\,\textrm{mm}$. 

We simulate the dipole distribution as two Gaussian patches: the first
one is narrow with the size of the atomic cloud in the z-wire trap
($\sigma_{\textrm{n},x}\approx170\,\mu\mathrm{m}$, $\sigma_{\textrm{n},y}\approx21\,\mu\mathrm{m}$),
and the second one is wide due to atoms expanding from the MOT ($\sigma_{\textrm{w}}=4.5\,\textrm{mm}$).
The size of the hole in the dipole distribution ($w_{\textrm{h}}=2\times\sigma_{\textrm{h}}=100\,\textrm{\ensuremath{\mu}m}$)
is taken to coincide with the waist of the blue laser beam. This choice
is obvious for the LIAD mechanism. For thermal desorption our model
(see \textit{Temperature simulation}) indicates that the temperature
profile is only slightly wider than $w_{\textrm{h}}$. We find that
for obtaining good fits at large duty cycles it is necessary to assume
that the hole size broadens. We describe this by introducing a saturation
mechanism. This saturation can be justified by the fact that the measured
electric field is very sensitive to the duty cycle for small duty
cycles, but loses its sensitivity for larger duty cycles. Also the
fact that the data for the case of the largest duty cycle is the least
noisy, suggests that the surface is then clean and the measurement
suffers less from shot to shot fluctuations of the adsorbant number. 

We describe the total dipole density as:

\begin{equation}
\rho_{\textrm{d}}=(\rho_{\textrm{w}}+\rho_{\textrm{n}})(1-h),
\end{equation}

\noindent where $h$ is the hole profile, $0<h<1$ with $h\rightarrow1$
corresponding to full depletion. In equation (3) $\rho_{\textrm{w}}$
is the broad truncated Gaussian distribution: 
\begin{align}
\rho_{\textrm{w}}(x,y) & =\begin{cases}
\frac{N_{\textrm{w}}}{2\pi\sigma_{\textrm{w}}^{2}}e^{-(x^{2}+y^{2})/2\sigma_{\textrm{w}}^{2}}, & |x,y|<L_{x,y}/2;\\
0, & \textrm{elsewhere};
\end{cases}
\end{align}
 and $N_{\textrm{w}}$ is the number of atoms in the Gaussian distribution.
The narrow Gaussian part is:

\begin{equation}
\rho_{\textrm{n}}(x,y)=\frac{N_{\textrm{n}}}{2\pi\sigma_{\textrm{n},x}\sigma_{\textrm{n},y}}e^{-x^{2}/2\sigma_{\textrm{n},x}^{2}-y^{2}/2\sigma_{\textrm{n},y}^{2}},
\end{equation}

\noindent with $N_{\textrm{n}}$ the number of atoms in the narrow
distribution. The hole profile is modeled as: 

\begin{equation}
h(x,y)=\frac{\gamma e^{-(x^{2}+y^{2})/2\sigma_{\textrm{h}}^{2}}}{1+\gamma e^{-(x^{2}+y^{2})/2\sigma_{\textrm{h}}^{2}}}.
\end{equation}
We introduce here a heuristic saturation parameter $\gamma$ used
as a fitting parameter in the model, as are $N_{\textrm{w}}$ and
$N_{\textrm{n}}$. This saturation model for the depletion provides
the necessary broadening mechanism for the dipole distribution. 

For the sake of stability of the fit, we set $\gamma=0$ for the $0.2\%$
duty cycle (blue open circles in Figure 2). This is justified by the
fact that this is the shortest duty cycle which still allows to see
the Rydberg depletion in the atomic cloud and is the best approximation
to the dipole distribution without the effect of the blue laser. Also
for stability reasons $N_{\textrm{n}}$ is set to $0$ for the highest
duty cycle data ($86\%$ duty cycle), justified by the fact that for
such a large saturation the small dipole distribution effectively
vanishes. All the curves are simultaneously fitted to a single set
of parameters: $N_{\textrm{w}}$ is the same for all six data sets,
$N_{\textrm{n}}$ is the same for all duty cycles except the highest
86\%, and$\gamma$ is individually fitted for each data set. The result
of the fit is shown in Figure 2. The error bars on the data are based
on the fit to the Stark maps. The curves are the result of the fit
of the hole in the dipole distribution. In the table on the right
hand side of Figure 2 for each curve the corresponding values of duty
cycle and saturation parameter are shown.

The fit gives the values for the dipole distribution: $N_{\textrm{w}}=(6.47\pm0.02)\times10^{13}$
dipoles and $N_{\textrm{n}}=(4.13\pm0.01)\times10^{9}$ dipoles. This
gives peak surface dipole density $\rho_{\textrm{max}}=7.03\times10^{5}\,\textrm{atoms}/\mu\mathrm{m^{2}}$.
The found dipole density gives a minimum average ad-atom spacing of
$\sim1.2\,\textrm{nm}$. This is of the same order as in other studies:
in \cite{Sedlacek2016} the estimated surface density and average
inter ad-atom spacing are $4\times10^{5}\,\textrm{atoms}/\mu\mathrm{m^{2}}$
and $\sim1.5\,\textrm{nm}$, respectively. The \textit{x} and \textit{y}
cuts of the dipole distributions for different values of $\gamma$
are shown in Figure 4. For the largest value of $\gamma=68$ (brown
curve) the saturation is clearly visible as a widening of the hole.

\subsubsection*{\[ \textrm{\textmd{\textit{b. Temperature simulation}}} \]}

\begin{figure}[t]
\includegraphics[clip,width=0.98\columnwidth]{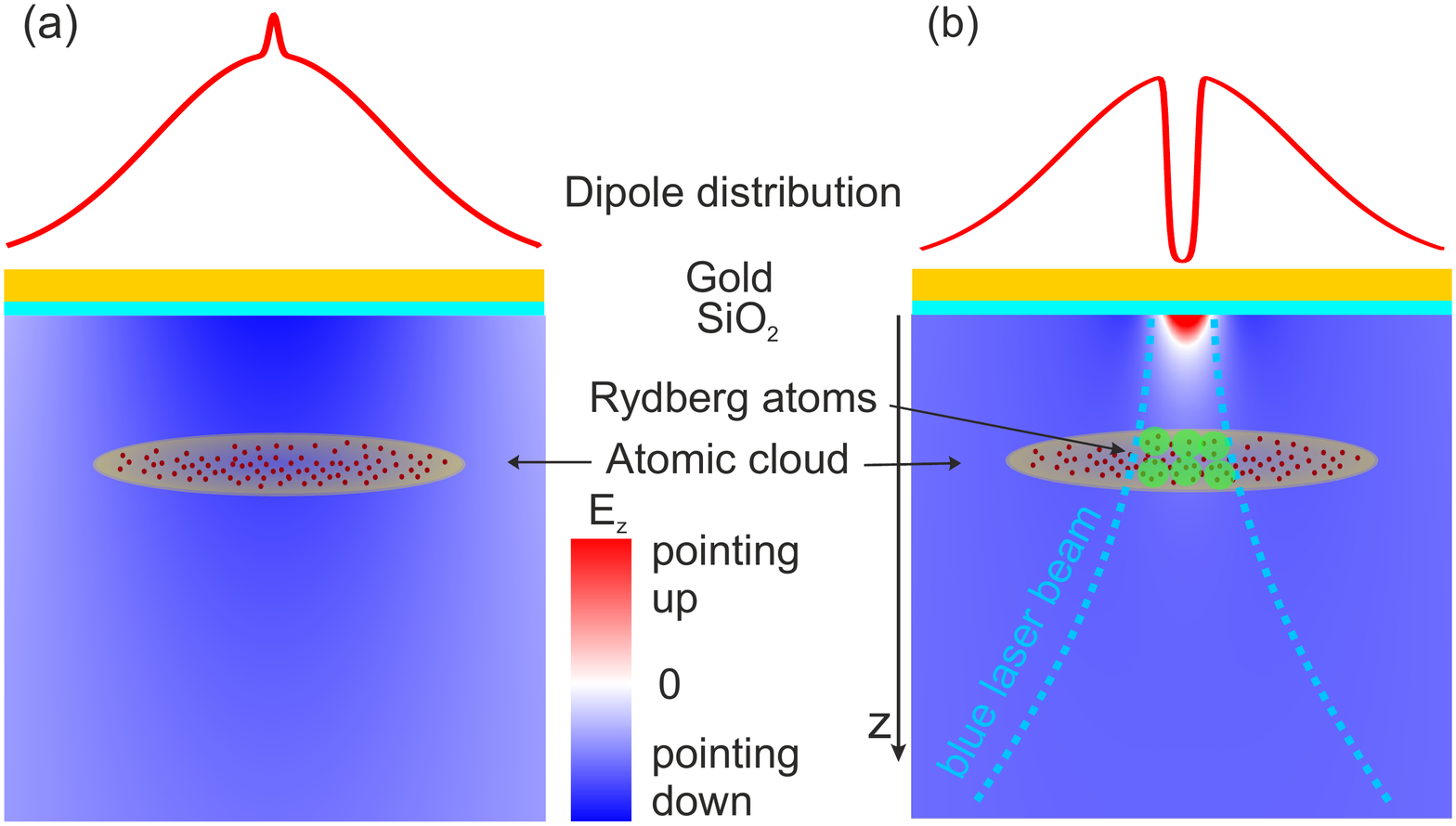}

{\footnotesize{}Figure 3. Simulation of the electric field $E_{z}$
due to a distribution of dipoles on the surface. The red curves show
the dipole distribution. The golden and the cyan parts show the two
outermost layers of the chip, the gold and the silica respectively.
The $E_{z}$ is indicated by color, blue (red) indicating electric
field pointing down (up). Figure (a) corresponds to a short duty cycle
of the blue laser (see }\textit{\footnotesize{}Electric field simulation}{\footnotesize{}),
(b) adsorbate distribution with a Gaussian hole in it, corresponding
to a long duty cycle of the blue laser. In the area where the blue
laser (dashed line) hits the surface, a hole in the adsorbate layer
occurs due to thermal and light-induced atomic desorption.}{\footnotesize \par}
\end{figure}

In order to understand the heating mechanism of the chip during the
experiment, the temperature of the chip was simulated. As shown in
Figure 1(c) the atom chip is a stack of layers with different materials.
The ad-atoms are adsorbed and desorbed from silica ($\sim25\,\mathrm{nm}$
layer). There is a thin layer of gold ($90\,\mathrm{nm}$) for the
mirror MOT, and a $1\,\mathrm{\mu m}$ layer of dielectric (SU8) for
planarization of the magnetic structure ($200\,\mathrm{nm}$ FePt),
which is used to create magnetic microtraps. The last layer of the
chip is a relatively thick ($\sim300\,\mathrm{\mu m}$) silicon substrate.
The chip is clipped to a metal construction, which is a good heat
conductor. We assume the connection of the chip to the metal construction
to be at room temperature $296\,\mathrm{K}$. The laser heats the
chip surface locally. The laser beam has a Gaussian profile with the
waist of $\sim100\,\mathrm{\mu m}$ and power of $55\,\mathrm{mW}$.
We estimate the absorption of the blue laser power by the chip to
be between $70\%$ and $80\%$. The heating problem can be solved
semi-analytically. We find that the center of the laser beam heats
up the chip surface by $\sim11-13\,\mathrm{K}$ above room temperature,
which leads to the enhanced desorption of the Rb atoms. The temperature
distribution closely follows the Gaussian form of the laser beam with
a root mean square width $\sigma_{T}\approx53\,\mathrm{\mu m}$, only
slightly larger than the laser beam $\sigma_{\textrm{laser}}=50\,\mathrm{\mu m}$.
The temperature profile resembles the laser beam profile almost perfectly
because the characteristic length scale of the temperature change
$r_{0}=\sqrt{Dd\lambda_{\textrm{Au}}/\lambda_{\textrm{SU8}}}\approx12\,\textrm{\ensuremath{\mu}m}$
is much smaller than the beam diameter, where $D$ and $d$ are the
thicknesses of the gold and SU8 layers respectively, $\lambda_{\textrm{Au}}=318\,\textrm{W/m\,K}$
and $\lambda_{\textrm{SU8}}=0.2\,\textrm{W/m\,K}$ are the thermal
conductivities.

The effective heating of the surface is due to the fact that gold
is a relatively bad mirror for the blue light, but mostly because
SU8 is a very good thermal insulator compared to gold. 

\begin{figure}[t]
\subfloat{\includegraphics[width=0.94\columnwidth]{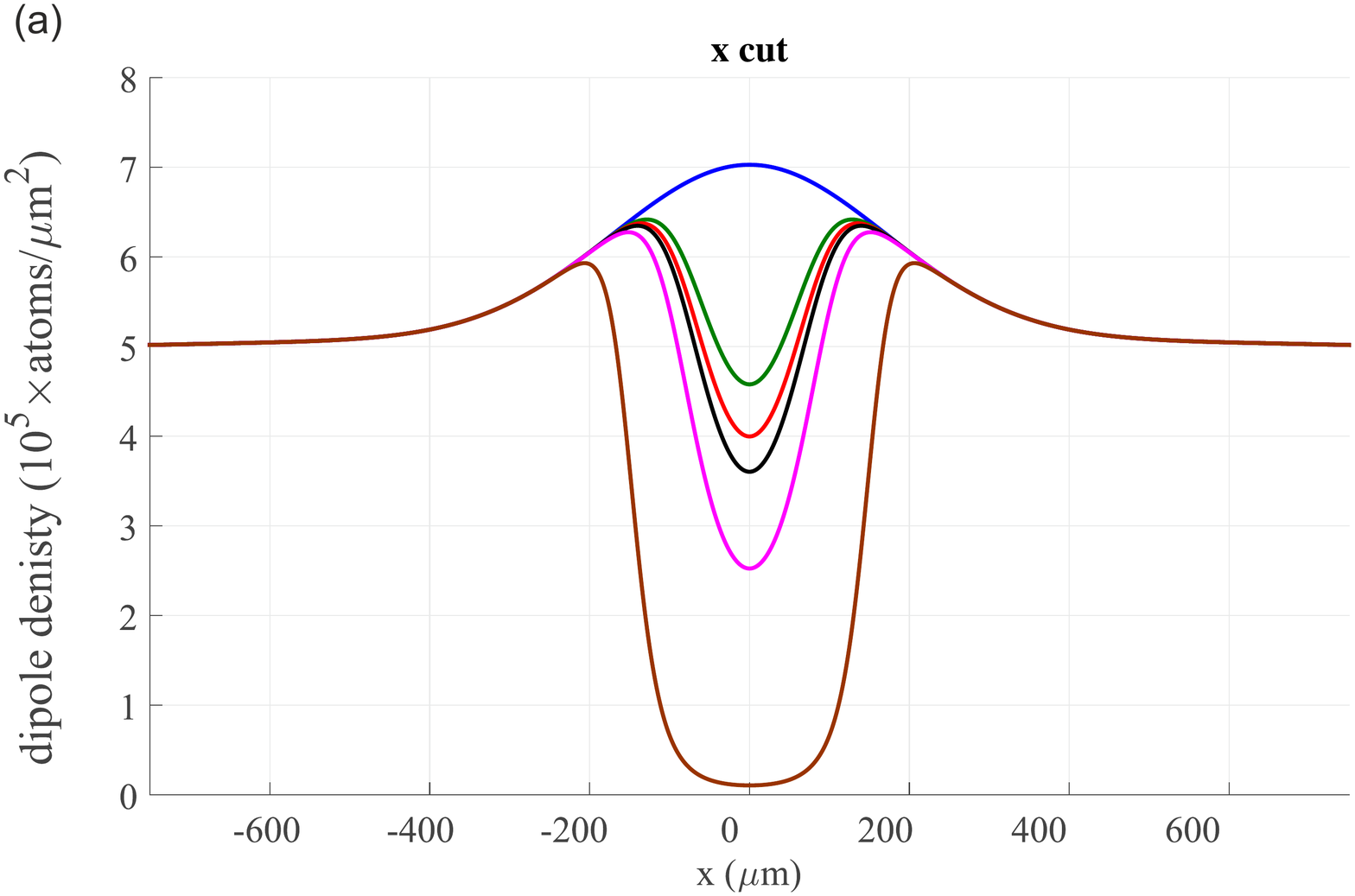}}

\subfloat{\includegraphics[width=0.94\columnwidth]{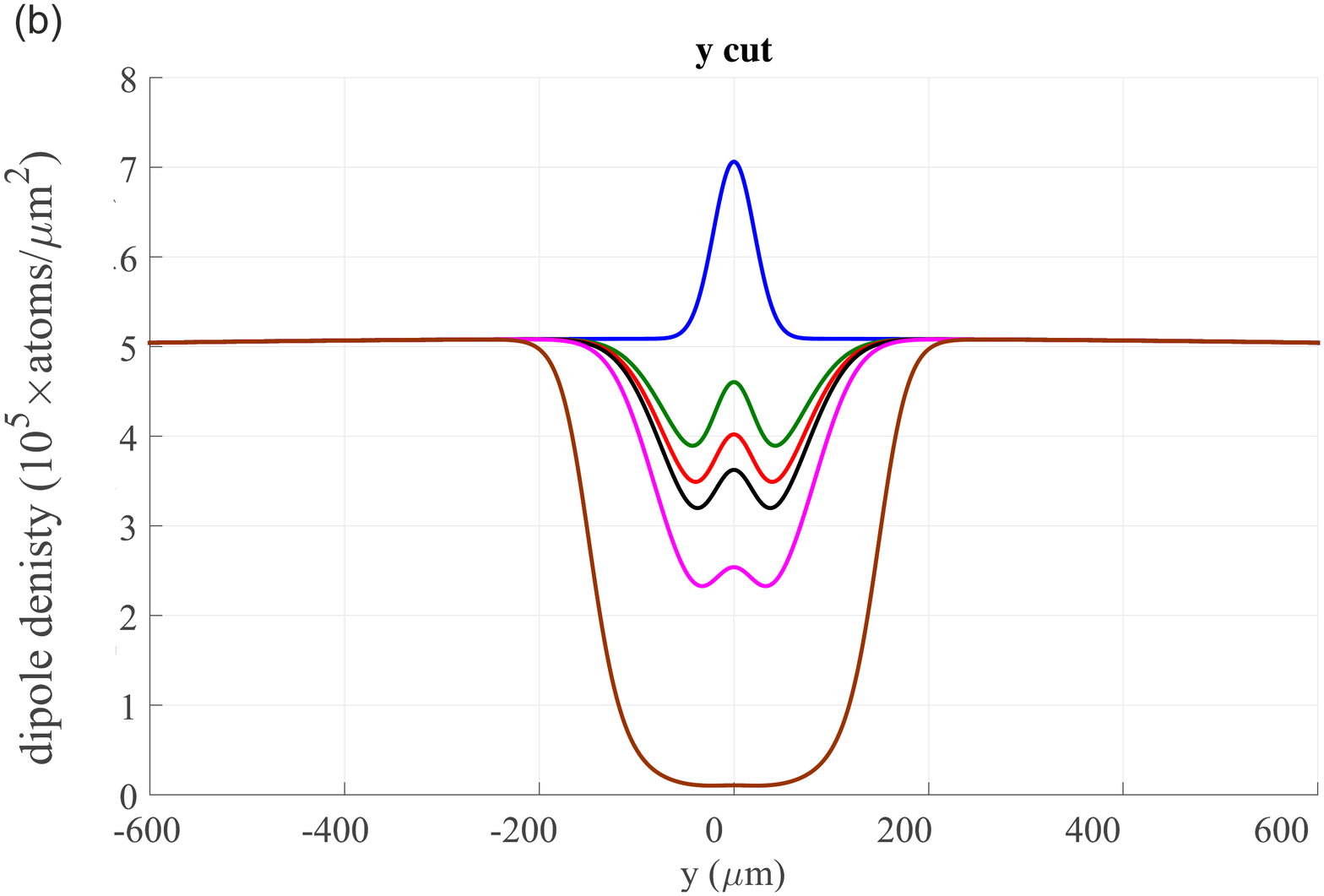}}

{\footnotesize{}Figure 4. X cut (a) and y cut (b) of dipole distributions
from the data fits for different duty cycles (see Figure 2). The distribution
consists of a 2D Gaussian on a broad layer of dipoles and a saturated
hole in it.}{\footnotesize \par}
\end{figure}

\subsubsection*{\[ \textrm{V. Discussion} \]}

By varying the average time of the blue laser acting on the chip we
are able to control the z-component of the stray electric field above
the chip surface. We attribute this effect to a combination of thermal
desorption and LIAD \cite{Burchianti2006}. Usually, for LIAD UV or
violet light is used \cite{Torralbo-Campo2015}, however in some special
cases of low adsorption energy the LIAD effect is seen even in the
red wavelength range \cite{Meucci1994}. In our case even though the
blue photon energy is not particularly high, the laser intensity is
several orders of magnitudes higher than in usual LIAD experiments.
Estimates for thermal desorption and LIAD based on the numbers provided
in \cite{Torralbo-Campo2015,Sedlacek2016} suggest that the desorption
is dominated by LIAD due to the high beam intensity, while the mild
local heating only contributes in a minor way. Both the precise adsorption-desorption
mechanism in our experiment and the dynamics of this process are subject
for future research.

Despite the strong electric field reduction in our experiment, no
reliable Rydberg excitations could be created in the microtraps at
$\sim10\,\textrm{\ensuremath{\mu}m}$ distance to the surface. As
the atomic cloud approaches the surface of the chip we see broadening
and suppression of the excitation spectra. This prevents Rydberg excitation
closer than $\sim40\,\mathrm{\mu m}$ from the chip, whereas the microtraps
are located $6-8\,\mathrm{\mu m}$ from the surface of the chip. As
mentioned before, one possible broadening mechanism is resonant dipole-dipole
interaction. Another source of spectral broadening and instability
of the stray electric field can be creation of free charges due to
the Rydberg ionization close to the surface and the deposition of
those charges onto the surface.

\subsubsection*{\[ \textrm{VI. Conclusion} \]}

We have demonstrated and analyzed a novel technique to control stray
electric fields in atom chip Rydberg experiments, changing the local
adsorbate dipole distribution by changing the surface conditions with
one of the two Rydberg excitation lasers. 

This effect of an excitation laser changing locally the surface dipole
distribution through heating and/or LIAD and thus changing the conditions
of the experiment therefore has to be taken into account in the design
of Rydberg experiments on a chip. 

The method presented in this paper provides a path forward towards
stable Rydberg excitation closer to the surface of an atom chip, in
particular in magnetic microtraps. The geometry of the affected area
can be further optimized by changing the beam parameters, for example,
by using a spatial light modulator \cite{Nogrette2014,Naber2016}.

\subsubsection*{\[ \textrm{Acknowledgments} \]}

Our work is financially supported by the Netherlands Organization
for Scientific Research (NWO). We also acknowledge financial support
by the EU H2020 FET Proactive project RySQ (640378).

\renewcommand\refname{\normalsize{}}

\bibliographystyle{unsrt}
\bibliography{MyCollection}

\end{document}